\RequirePackage{lineno}
\documentclass[reprint,aip,amsbsy,amsmath,amssymb,amsart,amsintx]{revtex4-1}

\usepackage{xcolor}
\usepackage{setspace}
\usepackage{graphicx}% Include figure files
\usepackage{epsfig}
\usepackage[latin1]{inputenc}
\usepackage[active]{srcltx}
\usepackage{dcolumn}% Align table columns on decimal point
\usepackage{bm}% bold math
\usepackage{epstopdf}

\begin{document}
\preprint{IST 3.2015-M J Pinheiro}

\title[]{Dimensional Analysis of Thrusting by Electromagnetic Inertia}

\author{Mario J. Pinheiro}
\address{Department of Physics, Instituto Superior Tecnico - IST, Universidade de Lisboa - UL, Av. Rovisco Pais, \& 1049-001 Lisboa, Portugal} \email{mpinheiro@tecnico.ulisboa.pt}

\thanks{We acknowledge financial support from Funda\c{c}\~{a}o para a Ci\^{e}ncia e Tecnologia through award SFRH/BSAB$/1420/2014$}


%\subjclass{}


\date{\today}
%\dedicatory{bla bla}%
%\commby{bla bla}
% ----------------------------------------------------------------
\begin{abstract}
We investigate the so called Biefeld-Brown effect in terms of
dimensional analysis and in the frame of classical
electrodynamics.
\end{abstract}
\maketitle

% ----------------------------------------------------------------
\section{Introduction}

The so called Biefeld-Brown effect is hypothetically a means to
convert electrostatic energy into a propulsive force, even in a
vacuum medium. It was discovered by Thomas Townsend Brown and Dr.
Paul Alfred Biefeld. In 1921 Brown discovered this effect when
experimenting with a Coolidge X-ray tube. The tube consisted of
two asymmetrical electrodes separated by a dielectric in a vacuum
environment. When connected to a high-voltage source a force acted
on the tube, forcing its motion in the direction of the positive
electrode~\cite{Brown1}. In case this effect is real the
associated potential for propulsion and as a source of energy is
enormous since no expenditure of fuel is necessary, the all
process remaining (most probably) nevertheless connected to
action-reaction type of momentum-transfer. Tests made at NASA
report thrust produced for various voltages, polarities and ground
configurations, leading to design asymmetrical capacitors for
propulsion~\cite{Canning}.

In 1893 Heaviside ~\cite{Heaviside1893} proposed the separation of
gravitation int electric and magnetic components.

Some Authors use the weak-field approximation to modify the
equations of the general theory of gravitation in a structurally
similar kind of equations as the Maxwell equations of the
electromagnetism ~\cite{Tajmar03}. We can find also in
Iwanaga~\cite{Iwanaga} a review of some field propulsion methods
based on general relativity theory. Nevertheless, so far proposed
concepts to control gravity for possible use on space propulsion
don't lead to no breakthrough ~\cite{Orfeu}.

Another interpretation relies on the standard atmospheric
electrodynamic model of the global electric circuit. The global
electric circuit is composed by the spherical conductor of the
earth, the spherical conductor of the ionosphere. Thunderstorms
act as charge generators, charging negatively the earth surface
and charging the ionosphere positive. Statistically the electric
field strength reaches maximums (at 18:00 UTC) and minimums (at
04:00 UTC) everywhere on earth according to UTC time and not the
local time. This was experimentally verified with data taken on
board the ship Carnegie and that is why is characteristic diurnal
curve is known as the "Carnegie curve". The self-potential
measured by Brown with a capacitor and a rock sample between the
electrodes acting as a dielectric, exposed to the Earth background
electric field apparently shows several parallels.
Stephenson~\cite{Stephenson} suggested that the Biefeld-Brown
effect is a secondary electrostatic effect related to the global
electric field. This interpretation precludes any relationship
with the electrogravitic nature of the effect.

Through a lagrangian formulation Feigel ~\cite{Feigel04} shown
that a body receives a recoil momentum from the vacuum equal to
the Minkowski's momentum, opening the possibility of contribution
of vacuum to the motion of dielectric liquids in crossed electric
and magnetic fields. Exploring different possibilities, Maclay and
Forward ~\cite{Forward2004} proposed a mechanism to propel a
spacecraft based on the Casimir effect, in which electromagnetic
radiation is emitted when an uncharged mirror is properly
accelerated in vacuum.

Following the claims from Podkletnov~\cite{Podkletnov} of the
possible connection between gravity and electro-magnetic effects
on type II, YBCO superconductors, experimental research has been
done to verify it without success~\cite{Robertson}.

Loder~\cite{Loder} gives an overall account of technologies
applications for the 21$^{st}$ century. In \cite{Puthoff} it is
discussed the possibility of engineering the zero-point field and
polarizable vacuum for propellantless propulsion and an idealized
system composed of two parallel semiconducting boundaries
separated by an empty gap variable width can, under appropriate
transformation, generate work~\cite{Pinto}.

\section{What is common knowledge about the Biefeld-Brown Effect}

According to experiments done by Thomas Townsend Brown which lead
to several patents registered in US and Great
Britain~\cite{Brown1,Brown2,Brown3}, the propulsive effect is
dependent on the following factors:
\begin{enumerate}
    \item the surface area of the electrodes, $S$;
    \item the voltage differential between the electrodes, $V$;
    \item the distance between the electrodes, $d$;
    \item the kind of material used between the masses, $\rho_m$;
    \item the dielectric permittivity of the material placed
    between the electrodes, $\epsilon_r$.
\end{enumerate}
This information suggests that the force is directly dependent on
the system capacitance and the total charge developed, $Q=CV$, as
the dimensional analysis of Section 2 clearly shows.

\section{A dimensional analysis}

We now apply the Buckingham's Pi theorem in order to obtain the
expected relationship between the operational variables related to
this problem, in particular, according to the Biefeld-Brown
investigations. In Table I we present the electromagnetic
magnitudes with relevance to our problem.

\begin{table}
  \centering
  \caption{Dimensions of the relevant physical variables}\label{Table1}
\begin{tabular}{|c|c|c|}
  \hline
  % after \\: \hline or \cline{col1-col2} \cline{col3-col4} ...
  Permittivity of vacuum & $[\varepsilon_0]$ = & $MLT^{-2}Q^{-1}$ \\
  \hline
  Electric scalar potential  & $[V]$ =& $ML^2T^{-2}Q^{-1}$ \\
  \hline
  Force & $[F]$ = & $MLT^{-2}$   \\
  \hline
\end{tabular}
\end{table}

Through a criterious application of the Buckingham's Pi theorem we
conclude that
\begin{equation}\label{Buck}
F=G \varepsilon \frac{V^2}{d} \mathcal{F} \left( \frac{S}{d}
\right)
\end{equation}
where $G$ is a constant to be determined by means of a theoretical
explanation or experimental data, and $\mathcal{F}(S/d)$ is a
function dependent on the ratio $S/d$.

\section{Interaction with the vacuum}

Although recently Newton's third law of motion was at
stack~\cite{Cornille95} it is likely action and reaction always
occurs by pairs and $\mathbf{F}=-\mathbf{F'}$ holds.

According to the Maxwell's theorem, the resultant of $\mathbf{K}$
forces applied to bodies situated within a closed surface $S$ is
given by the integral over the surface $S$ of the Maxwell
stresses:
\begin{equation}\label{Eq2}
\int \mathbf{T}(n) dS = \int \mathbf{f} d \Omega = \mathbf{K}.
\end{equation}
Here, $\mathbf{f}$ is the ponderomotive forces density and
$d\Omega$ is the volume element. The vector $\mathbf{T}(n)$ under
the integral in the left-hand side (lhs) of the equation is the
tension force acting on a surface element $dS$, with a normal
$\mathbf{n}$ directed toward the exterior. In cartesian
coordinates, each component of $\mathbf{T}(n)$ is defined by
\begin{equation}\label{}
T_x(n)=t_{xx} \cos (n,x) + t_{xy} \cos (n,y) + t_{xz} \cos (n,z),
\end{equation}
with similar expressions for $T_y$ and $T_z$. The 4-dimensional
momentum-energy tensor is a generalization of the 3-dimensional
stress tensor $T_{lm}$. If electric charges are inside a
conducting body in vacuum, in presence of electric $E$ and
magnetic $H$ fields, then Eq.~\ref{Eq2} must be modified to the
form
\begin{equation}\label{}
\int \mathbf{T}(n) dS - \mathbf{K} = \int \frac{1}{4 \pi c} \left(
\frac{\partial [\mathbf{E} \times \mathbf{H}]}{\partial t} \right)
d \Omega.
\end{equation}
In the right-hand side of the above equation it now appears the
temporal derivative of $\mathbf{G}=\int \mathbf{g}d \Omega$, the
electromagnetic momentum of the field in the entire volume
contained by the surface $S$ (with $\mathbf{g}$ its momentum
density).

In the case the surface $S$ is filled with a homogeneous medium
without true charges, Abraham proposed to write instead
\begin{equation}\label{Eq3}
\int \mathbf{T}(n) dS = \frac{\partial }{\partial t} \int
\frac{\varepsilon \mu}{4 \pi c} [\mathbf{E} \times \mathbf{H}] d
\Omega,
\end{equation}
with $\varepsilon$ and $\mu$ the dielectric constant of the medium
and its magnetic permeability.

Eq.~\ref{Eq3} can be written on the form of a general conservation
law
\begin{equation}\label{}
\frac{\partial \sigma_{\alpha \beta}}{\partial x_{\beta}} -
\frac{\partial g_{\alpha}}{\partial t} = f_{\alpha}
\end{equation}
where $\alpha=1,2,3$.

This equation can be reduced to the form
\begin{equation}\label{}
\frac{\partial \sigma_{\alpha \beta}}{\partial x_{\beta}}=
f_{\alpha}^L + \frac{1}{4 \pi c} \frac{\partial}{\partial t}
[\mathbf{D} \times  \mathbf{B}]_{\alpha} + f'_{m,\alpha}.
\end{equation}
Here, $f'_m$ is the force acting in the medium~\cite{Ginzburg76},
$\mathbf{f}^L=\rho_e \mathbf{E}+\frac{1}{c}[\mathbf{j} \times
\mathbf{B}]$ is the Lorentz force density with $\rho_e$ denoting
the charge density and $\mathbf{j}$ the current density.

Of course, the field and the the medium (or the matter) form
together a closed system and it is usual to catch the momentum
conservation law in the general form~\cite{Thirring,Landau2}
\begin{equation}\label{}
\frac{\partial (T_{\alpha \beta}^{Field} + T_{\alpha
\beta}^{Matter})}{\partial x_{\beta}} = 0.
\end{equation}

The general relation between Minkowski and Abraham momentum, free
of any particular assumption, holding particularly for a moving
medium, is given by
\begin{equation}\label{}
\mathbf{P}^M = \mathbf{P}^A + \int \mathbf{f}^A dt dV.
\end{equation}

For clearness, we shall distinguish between the parts of a system,
the body carrying currents and the currents themselves (the
structure for short), the fields and the vacuum.

The impulse transmitted to the structure is just
\begin{equation}\label{}
\mathbf{P}^K = \int \mathbf{f}^A dt dV = \mathbf{P}^M -
\mathbf{P}^A,
\end{equation}
where $\mathbf{f}^A$ is the Abraham's force density:
\begin{equation}\label{}
\mathbf{f}^A = \frac{\varepsilon_r \mu_r -1}{4 \pi c}
\frac{\partial [\mathbf{E} \times \mathbf{H}]}{\partial t}.
\end{equation}

This is in agreement with experimental data ~\cite{Jones} and was
proposed by others ~\cite{Gordon,Tangherlini}. As this force is
acting over the medium, it is expected nonlinearities related to
the behavior of the dielectric to different applied frequencies,
temperature,pressure, and large amplitudes of the electric field
when a pure dielectric response of the matter is no longer
proportional to the electric field (see Ref. ~\cite{Waser} on this
topic).

The momentum conservation law can be rewritten as (see
~\cite{Ginzburg76})
\begin{equation}\label{}
\frac{\partial \sigma_{\alpha \beta}}{\partial x_{\beta}} =
f_{\alpha}^L + \frac{1}{4 \pi c} \frac{\partial }{\partial t}
[\mathbf{D} \times \mathbf{B}]_{\alpha} + f_{m,\alpha}^{'},
\end{equation}
with $f_m^{'}$ denoting the force acting on the medium. The second
term in the lhs of above equation could possible be called
vacuum-interactance term ~\cite{Clevelance} - in fact, Minkowski
term. Already according to an interpretation of Einstein and Laub
~\cite{Laub}, the integration of above equation over all space,
the derivative over stress tensor gives a null integral and the
Lorentz forces summed over all the universe must be balanced by
the quantity $\int_{\infty} \varepsilon_0 \mu_0 \frac{\partial
[\mathbf{E} \times \mathbf{H}]}{\partial t} dV$ in order Newton's
third law be preserved.

As is well known, Maxwell's classical theory introduces the idea
of a real vacuum medium. After being considered useless by
Einstein's special theory of relativity, the ether (actually
replaced by the term {\it vacuum} or {\it physical vacuum}) was
rehabilitated by Einstein in 1920 ~\cite{Einstein22}. In fact,
general theory of relativity describes space with physical
properties by means of ten functions $g_{\mu \nu}$ (see also
~\cite{Ginzburg87}). According to Einstein, \begin{quote} The
ether of general relativity is a medium that by itself is devoid
of {\it all} mechanical and kinematic properties but at the same
time determines mechanical (and electromagnetic) processes.
\end{quote}

Dirac felt the necessity to introduce the idea of ether in quantum
mechanics ~\cite{Dirac}. In fact, according to quantum field
theory, particles can condense in vacuum giving rise to space-time
dependent macroscopic objects, for example, of ferromagnetic type.
Besides, stochastic electrodynamics shown that the vacuum contains
measurable energy called zero-point energy (ZPE) described as a
turbulent sea of randomly fluctuating electromagnetic field. Quite
interestingly, it was recently shown that the interaction of atoms
with ZPF guarantees the stability of matter and, in particular,
the energy radiated by an accelerated electron in circular motion
is balanced by the energy absorbed from the ZPF ~\cite{kozlowski}

Graham and Lahoz made three important
experiments~\cite{Lahoz1,Lahoz2,Lahoz3}. While the first
experiment provided an experimental observation of Abraham force
in a dielectric, the second one provided a measurement of a
reaction force which appear in magnetite. The third one provided
the first evidence of free electromagnetic angular momentum
created by quasistatic and independent electromagnetic fields $E$
and $B$ in the vacuum ~\cite{Note1}. Whereas the referred paper by
Lahoz provided experimental evidence for Abraham force at low
frequency fields, it still remains to gather evidence of its
validity at higher frequency domain, although some methods are
presently outlined~\cite{Antoci}.

All this is known since a long time and we only try to put more
clear the theoretical framework, that only needs to be
experimentally tested for proof of principles.

In view of the above, we will write the ponderomotive force
density acting on the composite body of arbitrarily large mass
(formed by the current configuration and its supporting structure)
in the form
\begin{equation}\label{}
\rho \frac{d \mathbf{V}}{dt} = \rho_c \mathbf{E} + [\mathbf{J}
\times \mathbf{B}] + \nabla \cdot \mathbf{T} +
\frac{\partial}{\partial t} \left( \varepsilon_0 \mu_0 [\mathbf{E}
\times \mathbf{H}] \right).
\end{equation}

Hence, the composite body is acted on by Minkowski force in such a
way that
\begin{equation}\label{}
M\mathbf{V} = -\mathbf{G}^M + \mathbf{G}^A.
\end{equation}
The Minkowski momentum is transferred only to the field in the
structure and not to the structure and the field in the
medium~\cite{Skobeltsyn,Ginzburg76,Lahoz3}.

It seems that generally in nature, locomotory propulsion by
oscillating flukes or wings are characterized by the periodic
shedding of wake vortices (the so called Von Karman streets)
inducing jet flows carrying momentum and thus the body will
experience a reaction force the propel it through the fluid
~\cite{Hendenstrom}. In a similar way it seems to occur with
Minkowski force, which is vortex formed in ether and inducing a
reaction force on the body which propels it in ether.

\section{Ponderomotive force via vector potential}

Difficulties related to the Minkowski-Abraham controversy maybe
avoided formulating the problem in term of canonical momentum,
instead of ponderomotive forces. Canonical momentum allows a clear
cut view of phenomena. In this theoretical frame there is no
violation of action-reaction law. In fact, the massless propulsion
is achieved obtaining mechanical momentum through electromagnetic
momentum exchanged with the medium.

According to Trammel ~\cite{Trammel}, the total momentum acting
over the composite body in an inertial frame can be also written
in the form
\begin{equation}\label{}
\mathbf{P} = \rho \mathbf{V} + \sum_i \rho_{c,i} \mathbf{v}_i +
\sum_i \rho_{c,i} \mathbf{A}_i.
\end{equation}
This is supposed to apply to a structure and current-carrying
wires integrated inside the structure. It is assumed to exist a
current density $\mathbf{J}=\sum_i \rho_{c,i} \mathbf{v}_i$. To
simplify, we will assume that in the structure inertial frame the
current undergoes arbitrary small acceleration, i.e., the current
is quasistationary. Also, we discard the effect of the motion on
the electromagnetic field.

We will assume a very simple geometry with two electrodes below
and above a dielectric cylinder (e.g, baryum titanate). The
leakage current will flow along z-axis and, consequently, so do
the potential vector $\mathbf{A}=A_z \mathbf{u}_z$. Hence, the
propulsive force will be acting along axis Oz:
\begin{equation}\label{}
\mathbf{F}_z = M \frac{d V_z}{d t} = - Q \frac{d A_z}{dt}
\end{equation}
where $M$ (an arbitrary large mass) and $V_z$ are the mass of the
composite body and its velocity along $z$-axis. The
electromagnetic part of momentum acting over the charge $Q$
developed on electrodes making part of the structure of mass $M$
is $QA_z$. $A_z$ is the vector potential component along Oz, given
by:
\begin{equation}\label{}
A_z (r) = \frac{\mu_0 J_z}{4} r^2.
\end{equation}
It is independent of z-coordinate, depending only on the radial
position. We denote by $J_z$ is the current density, $R$ the
electrodes radius (by default, the lower smaller radius) and $d$
the capacitors width. The total charge $Q$ present on each
electrode is such that $Q=C(V_1 - V_2)$.

After integration over the charged electrode, the final expression
for the force is readily obtained
\begin{equation}\label{Eq10}
F_z = - \frac{1}{8 c_0^2} \varepsilon_r \omega I \frac{R^2}{d}(V_1
- V_2).
\end{equation}
Our result working along a second line is consistent with Hector's
~\cite{Brito} finding, who obtained an analogous expression to
Eq.~\ref{Eq10}. In fact, Eq.~\ref{Eq10} is consistent with the
dimensional analysis done using Buckingham' Pi theorem, since a
dependency on energy retained between electrodes. If we restrain
ourselves to simple considerations, and as long as the leakage
current is given as $I\approx \Delta V/\mathcal{R}$, with
$\mathcal{R}$ representing now the electrical resistance of the
dielectric cylinder, the force is approximately given by
\begin{equation}\label{}
F_z \approx \frac{10^{-7}}{2} \frac{C \omega}{\mathcal{R}} \Delta
V^2.
\end{equation}
Eq.~\ref{Eq10} also is consistent with Eq.~\ref{Buck} obtained
through dimensional analysis, but now we have determined
explicitly the form of function $\mathcal{F}(S/d)$.

\section{Conclusion}

We investigate the so called Biefeld-Brown effect, or similar devices operating on electromagnetic forces, in terms of
dimensional analysis and in the frame of classical
electrodynamics. In particular, we obtained explicitly the function dependent form $\mathcal{F}$. This result can be helpful when addressing electromagnetic propulsion devices.

% ----------------------------------------------------------------
%\INPUT{Xbib.bib}   % For Gather Purpose Only
%\INPUT{Doc2.bbl}  % For Gather Purpose Only
\bibliographystyle{amsplain}
%\bibliography{Doc2}

\end{document}